\documentclass[aps,prl,twocolumn,superscriptaddress]{revtex4-2} % PRL style
\usepackage{amsmath,amssymb,graphicx,bm}
\usepackage{subcaption}
\usepackage[colorlinks=true,linkcolor=blue,citecolor=blue,urlcolor=blue]{hyperref}

\begin{document}

% Title and Authors
\title{\textit{C-BerryANC}: A \textit{first-principle} C++ code to calculate Berry Curvature dependent anomalous Nernst conductivity in any material}

\author{Vivek Pandey}
\altaffiliation{vivek6422763@gmail.com}
\affiliation{School of Physical Sciences, Indian Institute of Technology Mandi, Kamand - 175075, India}
\author{Sudhir K. Pandey}
\altaffiliation{sudhir@iitmandi.ac.in}
\affiliation{School of Mechanical and Materials Engineering, Indian Institute of Technology Mandi, Kamand - 175075, India}

\date{\today}

% Abstract
\begin{abstract}

The anomalous Nernst conductivity (ANC) is a key transport property in magnetic and topological materials, arising from the Berry curvature ($\boldsymbol\Omega$) of electronic bands and offering deep insight into their underlying topology and thermoelectric behavior. While Wannier interpolation methods have become popular for calculating ANC due to their computational efficiency, their accuracy critically depends on the quality of the Wannierization, which can be challenging for entangled bands or materials with complex band crossings. These limitations highlight the need for a direct \textit{first-principles} approach to reliably compute ANC from \textit{ab-initio} electronic structures. Here, we present a C++ based code named \textit{C-BerryANC} that calculates $\boldsymbol\Omega$-dependent ANC by directly using the eigenvalues and momentum-matrices obtained from DFT calculations. Presently, the code is interfaced with WIEN2k package which uses all-electron approach and full-potential linearized augmented plane wave (FP-LAPW) based method. For efficiently handling dense \textbf{\textit{k}}-mesh, calculation of $\boldsymbol\Omega$ is made parallel over \textbf{\textit{k}}-points using the OpenMP method. Additionally, the code stores band-resolved components of $\boldsymbol\Omega$ in binary files thereby reducing the memory occupancy and providing easy \& fast post-process option to compute ANC for any range of chemical potential and temperature values thus saving computational time. It is important to note that as compilation of C++ modules produce executable files which are in machine level language, computational speed of \textit{C-BerryANC} is very fast. The code is benchmarked over some well-known materials exhibiting ANC. These includes- Pd, Fe$_3$Al \& Co$_2$FeAl. The obtained values of ANC is found to in good agreement with the previously reported data. This highlights the accuracy, efficieny and reliability of the \textit{C-BerryANC} code.\\\\ 

% Berry curvature ($\boldsymbol\Omega$) associated with the Bloch states is known to influence several transport properties of any given material. Among these, some of the pronounced properties are anomalous Hall conductivity (AHC), anomalous Nernst conductivity (ANC), Berry curvature dipole and non-linear Hall effect. Conventional approach of calculating $\boldsymbol\Omega$ and related transport properties involves relying on wannerization techniques. However, this involves inaccuarcies, demanding the calculation of $\boldsymbol\Omega$-dependent properties using the \textit{first-principle} approach. In this regard, \textit{C-BerryANC} code is already designed to calculate $\boldsymbol\Omega$-dependent AHC using the WIEN2k output. In the present work, we extend the feature of \textit{C-BerryANC} code to much more advanced level. The present code in designed in C++ programming language. Thus, the post-processing task becomes much more faster (about 20 times) than \textit{C-BerryANC}. In addition to this, the code is capable of calculating other transport properties such as ANE, Berry curvature diploe ans non-linear Hall effect. Furthermore, unlike \textit{C-BerryANC} code which is only interfaced with the paid software WIEN2k, \textit{C-BerryTrans} is also interfaced with free DFT packages such as Elk and Abinit. \\\\
%

\textbf{Program summary -}\\
\textit{Program Title}: \textit{C-BerryANC}\\
\textit{Program Files doi}:\\
\textit{Licensing provisions}: GNU General Public License 3.0\\
\textit{Programming language}: C++\\
\textit{External routines/libraries}: Cmath, Vector, Chrono, Omp.h, Cstdlib, String\\
\textit{Nature of problem}: Computing ANC of any material using the \textit{first-principle} approach.\\
\textit{Solution method}: The code computes ANC by evaluating $\boldsymbol\Omega$ using the Kubo formula. It utilizes eigenvalues and momentum-matrix elements extracted directly from \textit{first-principles} DFT calculations performed with WIEN2k. The $\boldsymbol\Omega$ is then integrated over the full Brillouin zone (BZ), incorporating the entropy distribution to appropriately weight electronic states based on their occupation at various chemical potentials and temperatures.
\end{abstract}

\maketitle

% Main text
\section{Introduction}
Nernst effect is the observation of transverse voltage in a conducting material when it is subjected to a temperature gradient ($\boldsymbol\Delta\textbf{T}$) and magnetic filed \textbf{B}, both normal to each other\cite{behnia2016nernst,wang2006nernst}. This phenomena is important from thermoelectric point of view as it requires high electrical conductivity so as to conduct the charge particle and small thermal conductivity in order to maintain the temperature gradient\cite{mizuguchi2019energy,price1956theory}. This effect occur in ferromagnetic material even in the absence of external \textbf{B} and is proportional to magnetization\cite{lee2004anomalous,pu2008mott,sakuraba2013anomalous,hasegawa2015material}. The phenomena is termed as anomalous Nernst effect (ANE). With the recent discovery of Berry phase effects from Bloch states, it has been revealed that the Berry curvature ($\boldsymbol\Omega$) of the electronic band structure acts as intrinsic magnetic field which provides the mobile charge carriers a transverse velocity due to the Lorentz force\cite{liu2016quantum,shi2020anomalous,ikhlas2017large}. Thus, $\boldsymbol\Omega$ of the Bloch states give rise to ANE. This results to a transverse electrical conductivity popularly known as anomalous Nernst conductivity (ANC). Notably, this phenomena only arises in a material that breaks time-reversal symmetry leading to a net non-zero $\boldsymbol\Omega$.

ANE has an enormous practical applications. It is very efficient in thermal energy harvesting. In this direction, the effect has been recently found to produce high power density of around 13$\pm$2 W/cm$^3$ in nanofabricated device\cite{lopez2022high}. In addition to this, the phenomena is useful for designing heat flux sensors\cite{zhou2020heat}. The study of interplay between the spin Seebeck effect and ANE revealed the importance of ANE in spin-caloritronics\cite{huang2011intrinsic,bosu2012thermal,weiler2012local}. Also, ANE in a material enables its use in novel magnetic imaging microscopy techniques that rely on thermal gradients\cite{weiler2012local,reichlova2019imaging,janda2020magneto,johnson2022identifying}. Furthermore, ANE and its electrical counterpart \textit{i.e.}, anomalous Hall effect (AHE) is useful for the characterization of topological nature of charge carriers in magnetic Weyl semimetals. It it worth noting that unlike Anomalous Hall conductivity, which depends on overall integration of all the occupied bands, ANC arises only from the bands which contribute at the Fermi energy\cite{ikhlas2017large,xiao2006berry,xiao2010berry}. This makes the ANE measurement highly useful to characterize the $\boldsymbol\Omega$ distribution close to Fermi energy and to verify the possibility of Weyl phase. These discussion provides the motivation for efficient and accurate computation of ANC.

As suggested in the work of Xiao \textit{et. al.}, ANC ($\alpha_{ij}$) can be calculated using the relation\cite{PhysRevB.85.012405,xiao2010berry} 
\begin{multline}
\alpha_{ij} = -\frac{1}{T} \frac{e}{\hbar} \sum_n \int \frac{d^3k}{(2\pi)^3} \, \Omega_{ij}^n(\mathbf{k}) \left[ (E_n - E_F) f_n(\textbf{\textit{k}}) \right. \\
\left. + \, k_B T \ln \left(1 + \exp\left( \frac{E_n - E_F}{-k_B T} \right) \right) \right]
\label{eqANE}
\end{multline}
where, $f_n(\textbf{\textit{k}})$ is the Fermi-Dirac distribution function and $\Omega_{ij}^n(\textbf{\textit{k}})$ is the component of $\boldsymbol\Omega$ for band-index $n$ and at crystal momentum \textbf{\textit{k}}. Here, \textit{e}, $\hbar$, $k_B$ and $T$ denote the electronic charge, reduced Planck\textquoteright s constant, Rydberg constant and temperature, respectively. Considering the fact that $\Omega_{ij}^n(\textbf{\textit{k}})$ depends on momentum matrices and eigen-energies of Bloch states ($n$,\textbf{\textit{k}})\cite{chang1996berry,sundaram1999wave,onoda2002topological,jungwirth2002anomalous}, $\alpha_{ij}$ is highly sensitive to the details of electronic structure of any given material under study. Thus, the accuracy of $\alpha_{ij}$ depends on how accurately the Bloch states and the associated eigen-energies \& momentum-matrices are determined. 

Bloch states are extended in nature and are thus less preferred for exploring various electronic and dielectric properties of condensed matter system. Instead, the wannier representation of Bloch states is found to be more insightful for studying such properties\cite{kohn1959analytic,kohn1973wannier}. By wannier representation, it is meant that a set of localized wannier functions are constructed from the available Bloch states. Then, the Bloch states are expressed in the basis of the obtained wannier functions. Now, using the obtained Bloch states in terms of wannier functions, various properties of materials are efficiently explored. This procedure is popularly known as wannierization techniques\cite{koepernik2023symmetry}. However, due to the phase indeterminacy that the Bloch states $\psi_{n\textbf{k}}$ have at each \textbf{\textit{k}}-points, the corresponding wannier function is strongly non-unique in nature\cite{marzari2012maximally}. In such a scenario, one is interested in obtaining the maximally localized Wannier functions (MLWFs), as implemented in wannier90 package\cite{mostofi2008wannier90}. However, the procedure of doing so is highly parameter dependent and preferably demands the dispersion of material to follow particular features. The bands in the region of interest (around the Fermi energy in case of ANC) must be disentangled from the bands in other regions. This window is popularly known as disentangled energy window. The orbitals contributing to the bands in this window must have no or negligible contribution to the bands outside it. Furthermore, with the increase in dispersive nature of bands, Bloch states at more number of \textit{\textbf{k}}-points are required for obtaining MLWFs, which increases the computational cost. In this regard, it is worth noting that in certain cases (specially in metallic systems) where bands are highly dispersive, degenerate and extremely disentangled, obtaining MLWFs becomes a hopeless task. This demands developing of computational code that directly computes ANC at \textit{first-principle} level. To the best of our knowledge, such implementation is found to be missing in the existing family of codes and literature.

An efficient code is one which consumes minimum time and memory for accurate calculation of a given property. In the present context, ANC calculations using equation \ref{eqANE} require the integral of $\boldsymbol\Omega$ over an extremely densed \textbf{\textit{k}}-mesh of size 400$\times$400$\times$400 or more across the full BZ. Furthermore, a close inspection of the equation shows that if band-resolved $\boldsymbol\Omega$ at each \textbf{\textit{k}}-points is stored, then ANC can be computed at any combination of temperature and chemical potential value as post-process. Such a vast amount of data can be efficiently stored and handled in the binary file. This requires exceptionally small memory. In addition to this, the speed of computation can be enhanced by performing parallel calculations of band-resolved $\boldsymbol\Omega$ with respect to \textit{\textbf{k}}-points. In similar way, the post-process of ANC can also be made parallel. Designing a \textit{first-principle} code to compute ANC based on these approach is expected to be highly efficient. In addition to these aspects, the accuracy of ANC will depend on how accurately the momentum-matrices and eigenvalues associated with Bloch states are determined via DFT package. In the present scenario, DFT packages based on full potential linearized augmented plane wave method (FP-LAPW) is considered to be the most accurate one. Popular DFT packages that is based on this are WIEN2k\cite{blaha2020wien2k} and Elk\cite{ELK}. Additionally, WIEN2k is used by a vast research audience. Thus, designing a code to calculate ANC using WIEN2k output is expected to be highly accurate and beneficial to a large research community.

% In the present work, a C++ based code named \textit{C-BerryANC} is designed to compute $\boldsymbol\Omega$-dependent ANC using the \textit{first-principle} approach. 

In this work, we present a C++ based code named \textit{C-BerryANC}, designed to compute the $\boldsymbol\Omega$ dependent ANC directly from WIEN2k output. The code extracts eigenvalues and momentum-matrix elements from WIEN2k calculations and explicitly evaluates $\boldsymbol\Omega$ using a Kubo-like formula\cite{thouless1982quantized}. To efficiently handle large sets of \textbf{\textit{k}}-points, the $\boldsymbol\Omega$ calculations are parallelized across \textbf{\textit{k}}-space. To enhance the usability of the code, \textit{C-BerryANC} first computes band-resolved $\boldsymbol\Omega$ and stores it in a binary format, thereby greatly reducing memory usage. Subsequently, based on the temperature and chemical potential windows specified in the input file, ANC is calculated as a fast post-processing step. Since the $\boldsymbol\Omega$ evaluation constitutes the majority of the computational workload, this design allows for rapid evaluation of ANC over a wide range of temperatures and chemical potentials in a single run. Furthermore, \textit{C-BerryANC} provides a direct, interpolation-free approach that ensures high numerical accuracy and is well-suited for scalable computation. This feature is particularly beneficial in high-throughput materials screening, where manual tuning of Wannier functions becomes impractical. The code has been validated by benchmarking it against well-known test systems.

\begin{table*}
\caption{\label{tab:table1}%
\normalsize{Descriptions of the input parameters required for executing the \textit{C-BerryANC} code.}
}
\begin{ruledtabular}
\begin{tabular}{cc}
\textrm{\textbf{Name}} &
\textrm{\textbf{Meaning}} \\
\colrule
\textit{case} & Name of the WIEN2k directory containing self-consistently converged results with SOC enabled. \\
\textit{struct\_num} & Numerical identifier for the crystal structure type (see Table II for reference). \\
\textit{chemical\_potential} & Central value of the chemical potential (in Rydberg) used to define the energy window for ANC calculation. \\
\textit{emax} & Upper bound of the chemical potential window (in meV) relative to the specified \textit{chemical\_potential}. \\
\textit{emin} & Lower bound of the chemical potential window (in meV) relative to the specified \textit{chemical\_potential}. \\
\textit{estep} & Increment (in meV) for varying the chemical potential within the defined energy window. \\
\textit{Tmin} & Minimum temperature (in Kelvin) at which ANC will be evaluated. \\
\textit{Tmax} & Maximum temperature (in Kelvin) at which ANC will be evaluated. \\
\textit{Tstep} & Step size (in Kelvin) for varying temperature between \textit{Tmin} and \textit{Tmax}. \\
\textit{spin\_pol} (1/0) & Indicates whether spin-polarization is considered: 1 for spin-polarized systems, 0 otherwise.\\
                        &  SOC is always included when spin polarization is enabled. \\
\textit{kgrid} & Dimensions of the \textbf{\textit{k}}-mesh used for $\boldsymbol\Omega$ computations. \\
\textit{shift\_in\_k} (1/0) & Specifies whether the \textbf{\textit{k}}-mesh is shifted from high-symmetry points: 1 for shifted, 0 for unshifted mesh. \\
\textit{NUM\_THREADS} & Number of threads used to parallelize $\boldsymbol\Omega$ and ANC calculations across \textbf{\textit{k}}-points. \\
\textit{num\_kp\_at\_once} & Number of \textbf{\textit{k}}-points processed in a single batch per thread to manage memory usage efficiently and avoid overflow. \\

\textit{calc\_energy\_min} & Lower limit of the energy window (in meV and with respect to the specified \textit{chemical\_potential}) within which\\
                           & eigenvalues and momentum-matrices must be computed.\\
\textit{calc\_energy\_max} & Upper limit of the energy window (in meV and with respect to the specified \textit{chemical\_potential}) within which \\
                           & eigenvalues and momentum-matrices must be computed.\\
\end{tabular}
\end{ruledtabular}
\label{tab1}
\end{table*}

\begin{table}
\caption{\label{tab:table1}%
\normalsize{The structure number assigned to different crystal structures.
}}
\begin{ruledtabular}
\begin{tabular}{cc}
\textrm{\textbf{Crystal Structure}}&
\textrm{\textbf{Structure Number}}\\
\colrule
          Cubic Primitive           & 1\\
          Cubic face-centred        & 2\\
          Cubic body-centred        & 3\\
          Tetragonal Primitive      & 4\\
          Tetragonal body-centred   & 5\\
          Hexagonal Primitive       & 6\\
          Orthorhombic Primitive    & 7\\
          Orthorhombic base-centred & 8\\
          Orthorhombic body-centred & 9\\
          Orthorhombic face-centred & 10\\
          Trigonal Primitive        & 11 \\
          Hexagonal Primitive       & 12
    
\end{tabular}
\end{ruledtabular}
\label{tab2}
\end{table}

\section{Theoretical Background}

ANC arises from the $\boldsymbol\Omega$ of electronic bands in momentum space. $\Omega_{\xi}$ component of $\boldsymbol\Omega$ is related to an antisymmetric tensor $\Omega_{ij}^n$ via relation $\Omega_{ij}$ = $\epsilon_{ij\xi} \Omega_{\xi}$, where, $i$, $j$ \& $\xi$ represent $x$, $y$ \& $z$ cartesian direction as per required by the expression. Moving further, $\Omega_{ij}$ can be computed by substituting the eigenvalues and momentum-matrices obtained from WIEN2k calculations into the Kubo-like formula\cite{stejskal2023theoretical}:
\begin{equation}
   \Omega_{ij}^n(\textbf{\textit{k}})=-\frac{\hbar^2}{m^2} \sum_{n^{\prime} \neq n} \frac{2 \operatorname{Im}\left\langle\psi_{n \textbf{\textit{k}}}\right| p_i\left|\psi_{n^{\prime} \textbf{\textit{k}}}\right\rangle\left\langle\psi_{n^{\prime} \textbf{\textit{k}}}\right| p_j\left|\psi_{n \textbf{\textit{k}}}\right\rangle}{\left(E_n-E_{n^{\prime}}\right)^2}
   \label{eq2}
\end{equation}
where $E_n$ is the band-energy, $\psi_{n \textbf{\textit{k}}}$ are the Bloch states and $p_i$ are the momentum operators. Here, $\hbar$ \& $m$ denote the reduced Planck\textquoteright s constant and mass of free electron, respectively. Having obtained $\Omega_{\xi}$, ANC ($\alpha_{ij}$) can be computed using equation \ref{eqANE}. The presence of $f$ \& $T$ in the equation allows the computation of ANC at various combination of chemical potential and temperature. Additionally, it is well-known fact that to obtain a non-vanishing value of integral in equation \ref{eqANE}, one must necessarily take spin–orbit coupling (SOC) into account\cite{vanderbilt2018berry}. Thus, \textit{C-BerryANC} intrinsically takes SOC into account.

% The momentum-matrix elements ($\textless \psi_{n\textbf{k}}| \textbf{\textit{p}}| \psi_{m\textbf{k}} \textgreater$) required for $\boldsymbol\Omega$ calculations can be obtained using WIEN2k's OPTIC module. Here, $\textbf{\textit{p}}$ is the momentum operator. The workflow for extracting $\boldsymbol\Omega$ involves- (I) Perform self-consistent SOC included DFT calculations to obtain ground state charge density, (II) obtain the eigenvalues at a given \textbf{\textit{k}}-point, (III) obtain the momentum-matrix elements at the given \textbf{\textit{k}}-point, and (IV) calculate $\Omega_{ij}^n(\textbf{\textit{k}})$ at the given \textbf{\textit{k}}-point using equation \ref{eq2}. Having obtained the $\boldsymbol\Omega$, the integration in equation \ref{eqANE} is approximated by summing the values of $\Omega_{ij}^n(\mathbf{k}) \left[ (E_n - E_F) f_n(\textbf{\textit{k}}) \right. \\
% \left. + \, k_B T \ln \left(1 + \exp\left( \frac{E_n - E_F}{-k_B T} \right) \right) \right]$ over a dense \textbf{\textit{k}}-mesh uniformly distributed across the BZ.

$\boldsymbol\Omega$ is computed using the momentum matrix elements $\langle \psi_{n{\textit{\textbf{k}}}} | {\textit{\textbf{p}}} | \psi_{m{\textit{\textbf{k}}}} \rangle$, where ${\textit{\textbf{p}}}$ represents the momentum operator. These elements are extracted from the WIEN2k output via its \texttt{OPTIC} module. The calculation sequence begins with a self-consistent DFT run including SOC to obtain the ground-state charge density. Following this, the energy eigenvalues and momentum matrix elements are obtained at each ${\textit{\textit{k}}}$-point. Using this information, $\Omega_{ij}^n({\textbf{k}})$ is using equation \ref{eq2}. Once the $\boldsymbol\Omega$ is known, the ANC is determined by summing the integrand of equation \ref{eqANE} across a dense and uniform ${\textit{\textbf{k}}}$-mesh spanning the BZ. The integrand consists of $\Omega_{ij}^n({\textbf{k}})$ weighted by the entropy-related kernel:
\[
(E_n - E_F) f_n({\textit{\textbf{k}}}) + k_B T \ln \left(1 + \exp\left( \frac{E_n - E_F}{-k_B T} \right) \right)
\]
This summation numerically approximates the full BZ integration needed for evaluating the temperature- and chemical-potential-dependent ANC.

\section{Workﬂow and technical details}

\subsection{Workﬂow}

% \begin{figure}[t]
%     \centering
%     \includegraphics[width=0.50\textwidth,height=14.0cm]{fig1.eps}
%     \caption{Workﬂow of the \textit{C-BerryANC} code.}
%     \label{workflow}
% \end{figure}

For using the \textit{C-BerryANC} code, user needs to install it by running the \textit{make} command. This will produce the executable files corresponding to the ten modules of \textit{C-BerryANC}. Having done this, a \textit{bin} folder will be created in which all these executable files will be moved. The user is now required to define the path of the \textit{bin} folder in \textit{.bashrc} file. With this, the installation procedure is complete. The detailed working of the code is described below.

%%%%%%%%%%%%%%%%%%%%%%%%%%%%%%%%%%%
For performing the ANC calculations, user needs to first prepare a WIEN2k \textit{case} directory that contains self-consistently converged ground-state files with SOC included. Nextly, the input file for \textit{C-BerryANC} code, the \textit{C-BerryANC.input} must be prepared. In this file, the name of folder containing self-consistently converged ground-state files must be assigned to \textit{case} variable. The structure number of material under study must be given to \textit{struct\_num} variable. In this regard, table II defines the structure number assigned to various crystal structures in \textit{C-BerryANC} code. Furthermore, it is essential to specify whether the material under investigation is spin-polarized by assigning a value of 1 (for spin-polarized) or 0 (for non-spin-polarized) to the \textit{spin\_pol} variable. Moving next, it is commonly observed that WIEN2k takes exceptionally long time to produce \textit{case.klist} file for a dense \textit{\textbf{k}}-mesh size (for instance 400$\times$400$\times$400). To ease this, \textit{C-BerryANC} is provided with a separate module named \textit{klist\_gen\_div}. Thus, if one needs to calculate ANC on 400$\times$400$\times$400 \textbf{\textit{k}}-mesh size, then the string \textquoteleft 400 400 400\textquoteright\hspace*{0.02in} must be assigned to \textit{kmesh} variable in the \textit{C-BerryANC.input} file. Using this information and the value assigned to \textit{struct\_num} variable, \textit{klist\_gen\_div} module efficiently produces the \textit{case.klist} file. In this regard, to obtain the shifted \textit{\textbf{k}}-mesh (that is, none of the \textbf{\textit{k}}-points of the mesh should correspond to a high-symmetric point), 1 must be assigned to variable \textit{shift\_in\_k}, otherwise it must be assigned 0. The \textit{case.klist} file obtained from \textit{klist\_gen\_div} is exactly the same as the one which WIEN2k produces for the given \textbf{\textit{k}}-mesh size. Moving further, the calculation of ANC via \textit{C-BerryANC} can be performed in serial or parallel mode depending on whether \textit{NUM\_THREADS} variable is assigned value 1 or greater than 1, respectively. If \textit{NUM\_THREADS}$>$1, then \textit{klist\_gen\_div} divides the generated \textit{case.klist} file into \textit{NUM\_THREADS} parts with names \textit{case.klist1}, \textit{case.klist2}, ..., \textit{case.klistNUM\_THREADS}. The total number of \textbf{\textit{k}}-points in \textit{case.klist} file is equally distributed among these files. Nextly, \textit{C-BerryANC} can compute ANC for a series of temperature values in a given temperature window at a time. For this, user needs to provide the lower-limit, upper-limit and the temperature step to variable \textit{Tmin}, \textit{Tmax} and \textit{Tstep}, respectively. All these values must be given in units of Kelvin. At the same time, for each temperature value, ANC can be computed for a series of chemical potential values within a given energy window range. For this, user needs to firstly assign the chemical potential value to the variable \textit{chemical\_potential}. Then, the lower-range (in meV and with respect of the value assigned to \textit{chemical\_potential}), upper-range (in meV and with respect of the value assigned to \textit{chemical\_potential}) and chemical potential step-size (in meV) must be assigned to variable \textit{emin}, \textit{emax} and \textit{estep}, respectively. It is important to highlight here that while calculating the eigenvalues via WIEN2k, the memory occupied in RAM increases with the increase in number of \textbf{\textit{k}}-points in \textit{case.klist} file. Thus, depending on the size of RAM, beyond a given size of \textbf{\textit{k}}-mesh, RAM gets fully occupied and the calculations get aborted. To avoid this situation, \textit{C-BerryANC.input} file is provided with a variable named \textit{num\_kp\_at\_once}. The value of this variable decides the number of \textbf{\textit{k}}-points over which $\boldsymbol\Omega$ calculation is carried out at each thread at any given time. Lastly, as already discussed, ANC depends only on the states around the Fermi energy. Thus, for computing the integration in equation \ref{eqANE}, the band-energy and momentum-matrices for bands far from the Fermi energy must not be computed. This will save the computation time. In this regard, user is allowed to provide the energy window corresponding to which energy and momentum-matrices must be calculated. The lower-range and upper-range of this window (in meV and with respect to value provided to \textit{chemical\_potential} variable) must be provided to variables \textit{calc\_energy\_min} \& \textit{calc\_energy\_max} in the input file, respectively. Based on these values, \textit{C-BerryANC} will automatically modify the \textit{case.in1}, \textit{case.inso} and \textit{case.inop} file to enhance the speed of computation. The details of the changes in these files that the code performs can be found in reference \cite{pandey2025textit}. With this, the discussion about various input variables of \textit{C-BerryANC} code is done. Now, a brief discussion about the workflow of the code is mentioned below.

\begin{figure}[t]
    \centering
    \includegraphics[width=0.50\textwidth,height=14.0cm]{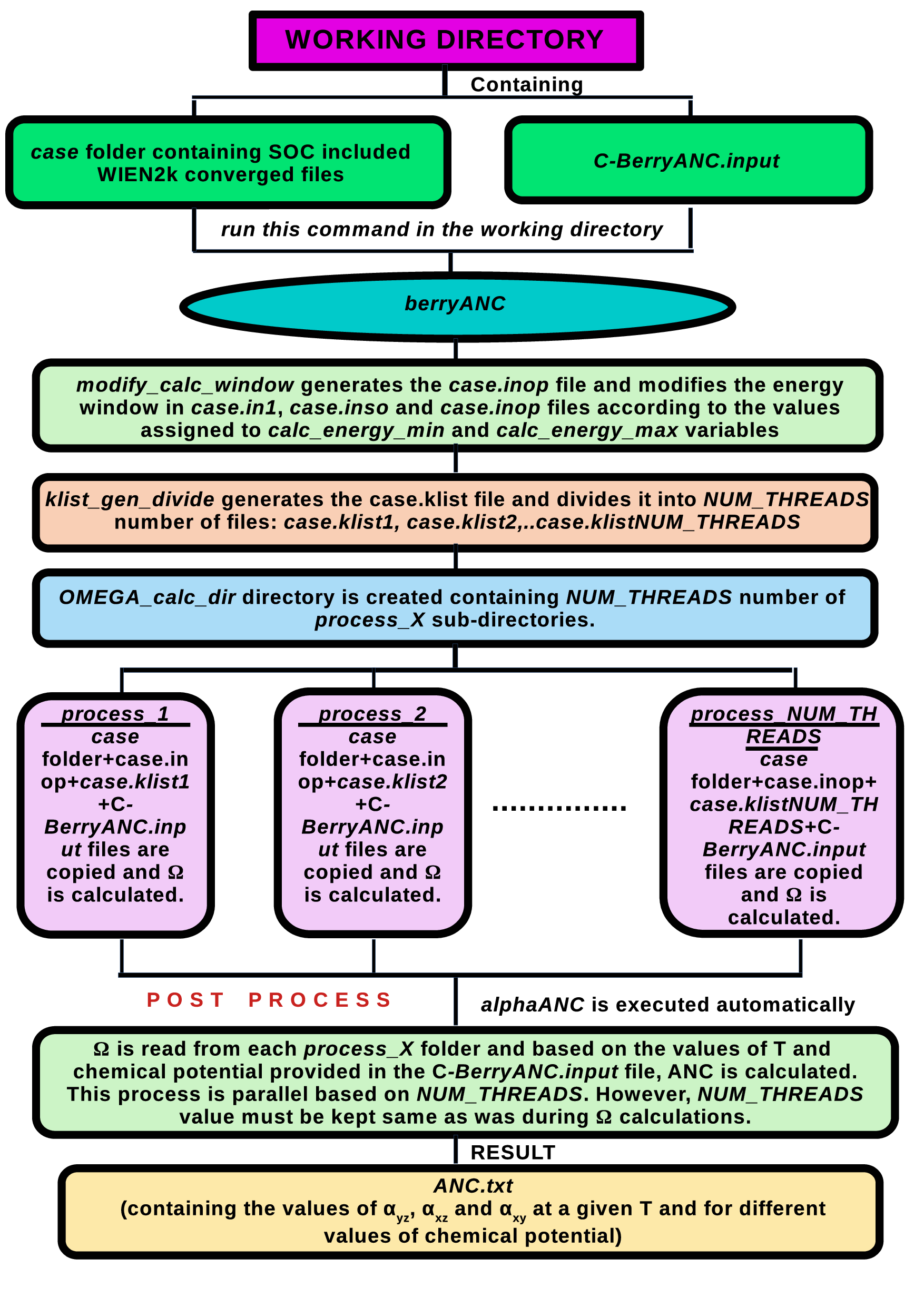}
    \caption{Workﬂow of the \textit{C-BerryANC} code.}
    \label{workflow}
\end{figure}

The flowchart of the working of \textit{C-BerryANC} code is shown in figure \ref{workflow}. Firstly, the code calls \textit{modify\_calc\_window} module to produce \textit{case.inop} file. After this, the module modifies the energy window in \textit{case.in1}, \textit{case.inso} and \textit{case.inop} files as per the values provided to the variables \textit{calc\_energy\_min} \& \textit{calc\_energy\_max} in the input file. Having done this, the code executes the \textit{klist\_gen\_div} module to generate the \textit{case.klist} file based on the \textbf{\textit{k}}-mesh size assigned to \textit{kmesh} variable. Nextly, if \textit{NUM\_THREADS}$>$1, the \textbf{\textit{k}}-points in \textit{case.klist} file is equally divided into \textit{NUM\_THREADS} number of files as described above. Now, a directory named \textit{OMEGA\_calc\_dir} is created within which \textit{NUM\_THREADS} number of subdirectories a created with names- \textit{process\_1}, \textit{process\_2}, ..., \textit{process\_NUM\_THREADS}. In each of these subdirectories, the \textit{case} folder along with \textit{case.inop} and \textit{case.klistN} (where \textit{N} is the number associated with the subdirectory) files are copied. The \textit{case.klistN} file is renamed as \textit{case.klist1} after being copied in the subdirectories. Now, the calculation of eigenvalues and momentum-matrices along with the computation of $\boldsymbol\Omega$ corresponding to \textit{case.klist1} file in each of these subdirectories will be carried out at separate threads. In this way the computation of $\boldsymbol\Omega$ is made parallel. As highlighted above, to avoid the over usage of RAM, in each of the subdirectories, firstly \textit{num\_kp\_at\_once} number of \textbf{\textit{k}}-points will be taken out of \textit{case.klist1} file to make a new file named \textit{case.klist}. Now the calculation of eigen energy and momentum-matrices corresponding to the generated \textit{case.klist} file will be done. After this, the band-resolved $\boldsymbol\Omega$ will be calculated for the \textbf{\textit{k}}-points in \textit{case.klist} file by calling the module \textit{energy\_pmat\_read}. The obtained result will be stored in binary file named \textit{data.bin}. Again, the next \textit{num\_kp\_at\_once} number of \textbf{\textit{k}}-points will be taken out of \textit{case.klist1} and the process will be repeated till the computation of band-resolved $\boldsymbol\Omega$ is done for all the \textbf{\textit{k}}-points in \textit{case.klist1} file. After this process is complete, everything is set for the post-process (calculation of $\alpha_{ij}$).

The post-process is accomplished by calling the module \textit{alphaANC} which computes the value of $\Omega_{ij}^n(\textbf{\textit{k}}) \left[ (E_n - E_F) f_n(\textbf{\textit{k}}) + k_B T \ln \left(1 + \exp\left( \frac{E_n - E_F}{-k_B T} \right) \right) \right]$ for all the bands at each \textbf{\textit{k}}-points. In a single run, it performs this calculations for a range of temperature and chemical potential values. For this, the temperature range is decided by the values assigned to variables \textit{Tmin}, \textit{Tmax} and \textit{Tstep} while the chemical potential window is given by the values provided to variables \textit{emin}, \textit{emax} and \textit{estep}. Finally, the module calculates $\alpha_{ij}$ (for $ij$= {$xz$, $yz$ \& $xy$}) for each combination of temperature and chemical potential by summing up the respective values of $\Omega_{ij}^n(\textbf{\textit{k}}) \left[ (E_n - E_F) f_n(\textbf{\textit{k}}) + k_B T \ln \left(1 + \exp\left( \frac{E_n - E_F}{-k_B T} \right) \right) \right]$ for all the bands and across all the \textit{\textbf{k}}-points of the BZ. The results obtained is mentioned in \textit{ANC.txt} file. It is important to note that while running the \textit{C-BerryANC} code, the post-process is automatically carried out after the calculation of $\boldsymbol\Omega$. However, once the \textit{data.bin} files are created after the calculations of $\boldsymbol\Omega$, one has freedom to run post-process (by calling the command \textit{alphaANC}) for any range of chemical-potential and temperature values by doing the respective modification in the \textit{C-BerryANC.input} file. It is important to note that the post-processing should be performed using the same number of threads as those used during the $\boldsymbol\Omega$ calculations. Furthermore, total time taken for $\boldsymbol\Omega$ calculation is mentioned in \textit{time.dat file}.

\subsection{Technical details}
All the modules of \textit{C-BerryANC} code is written in C++. Various packages of C++ like- \textit{cmath}, \textit{vector}, \textit{chrono}, \textit{omp.h}, \textit{cstdlib}, \textit{string}, etc have been explicitly used in it. Moving next, the calculation of $\boldsymbol\Omega$ (and ANC in the post-process) is made parallel over the \textbf{\textit{k}}-points using the OpenMP method. In addition to this, computation of $\boldsymbol\Omega$ \& ANC is done using the \textit{ab-initio} approach. Presently, the code is interfaced with the WIEN2k package. However, it can be easily interfaced with the other DFT packages like- ELK\cite{ELK}, Quantum Expresso\cite{giannozzi2009quantum}, \textit{etc}. Also, the code is implemented in a fully general framework, without assumptions tied to any particular material class or model, ensuring broad applicability and practical versatility across a wide range of systems. 

It is well-known that modules written in C++, when compiled produces executable files. Since these executables are already written in machine level language, they are comparatively vary fast than modules written in programming languages like python. In addition to this, unlike \textit{multiprocessing} class of python 3, the \textit{OpenMP} method in C++ is much more efficient in terms of computational time. Thus, \textit{C-BerryANC} is expected to provide ANC results in minimal computational time. \\\\

% \textbf{\underline{Strategy for efficient use of the code}:} A close inspection of equations \ref{eqANE} \& \ref{eq2} suggest that for a given value of $T$, a pair of bands will not contribute to AHC if at each \textbf{\textit{k}}-points of the BZ, the occupancy of one band is exactly equal to the other band. The pair of bands will only have non-zero contribution to total AHC at the given value of $T$, if in some regions of the BZ, the occupancy of one of the band differs from the other. Next thing that these equations suggest is that if a pair of bands have large energy difference, then their contributions to AHC will be negligible and thus can be safely ignored in computations. This discussion suggests that calculations of all the bands at a given \textbf{\textit{k}}-points, which generally consumes higher computational resources is not really required. It is therefore suggested that for upto temperature of 300 K, AHC at a given chemical potential ($\mu$) can be computed by taking the energy interval of not more than $\pm$1.5 eV with respect to $\mu$. The procedure to achieve this in WIEN2k calculations is discussed further.

\begin{table*}
\caption{\label{tab:table1}%
\normalsize{The various input details and the \textit{\textbf{k}}-mesh size used to calculate the ground state energy of the materials using WIEN2k package. The \textit{\textbf{k}}-meshes are taken in the irreducible part of the Brillouin zone (IBZ).
}}
\begin{ruledtabular}
\begin{tabular}{cccccc}
\textrm{$\textbf{case}$}&
\textrm{$\textbf{Space-Group}$}&
\textrm{$\textbf{Lattice Parameters}$}&
\textrm{$\textbf{Wyckoff Positions}$}&
\textrm{\textbf{\textit{k}-mesh (in IBZ)}}&
\textrm{$\textbf{Exchange–Correlation}$}\\
\colrule\\
    Pd\cite{guo2014anomalous}   &$Fm\bar{3}m$   & a=b=c=3.89 \AA      & Pd = (0.00, 0.00, 0.00)             &  10$\times$10$\times$10 & PBE \\
                     &          & $\alpha$=$\beta$=$\gamma$=90       &           &                       \\\\
Fe$_3$Al\cite{nishino1997semiconductorlike} &$Fm\bar{3}m$ & a=b=5.77 \AA       & Fe (I) = (0.00,0.00,0.50)        & 10$\times$10$\times$10 & PBESol \\
         &           & $\alpha$=$\beta$=$\gamma$ 90 & Fe (II) = (0.75, 0.25, 0.75)        &                         \\
         &           &                                    & Al = (0.00, 0.00, 0.00)              &                         \\\\
Co$_2$FeAl\cite{shukla2022atomic} &$Fm\bar{3}m$ & a=b=c=5.70 \AA              & Co = (0.25, 0.25, 0.25)        & 10$\times$10$\times$10  & PBESol \\
         &           & $\alpha$=$\beta$=$\gamma$= 90     & Fe = (0.50, 0.50, 0.50)        &                         \\
         &           &                                   & Al = (0.00, 0.00, 0.00)              &                         \\\\
\end{tabular}
\end{ruledtabular}
\label{tab3}
\end{table*}

\begin{table*}
\caption{\label{tab:table1}%
\normalsize{The details of the values assigned to the input parameters for the \textit{C-BerryANC} code. The \textbf{\textit{k}}-grid is sampled across the full Brillouin zone (FBZ)..
}}
\begin{ruledtabular}
\begin{tabular}{cccccc}
\textrm{$\textbf{\textit{case}}$}&
\textrm{$\textbf{\textit{struct\_num}}$}&
\textrm{$\textbf{\textit{chemical\_potential}}$}&
\textrm{$\textbf{\textit{spin\_pol}}$}&
\textrm{$\textbf{\textit{kgrid}}$}&
\textrm{$\textbf{\textit{shift\_in\_k}}$}\\
\textrm{$\textbf{ }$}&
\textrm{$\textbf{ }$}&
\textrm{$\textbf{(Rydberg)}$}&
\textrm{$\textbf{ }$}&
\textrm{$\textbf{ }$}&
\textrm{$\textbf{ }$}\\
\colrule
    Pd     & 2 &  0.5153375149 & 1 & 400$\times$400$\times$400 & 1\\
Fe$_3$Al   & 2 &  0.6017577882 & 1 & 400$\times$400$\times$400 & 1\\
Co$_2$FeAl & 2 &  0.6276267143 & 1 & 400$\times$400$\times$400 & 1\\
\end{tabular}
\end{ruledtabular}
\label{tab31}
\end{table*}

\section{Test Cases}
\textit{C-BerryANC} has been benchmarked over three ferromagnetic materials: Pd\cite{guo2014anomalous}, Fe$_3$Al\cite{nishino1997semiconductorlike} and Co$_2$FeAl\cite{shukla2022atomic}. All the three materials crystallize in face-centered cubic crystal structure. These material are expected to exhibit ANE due to their intrinsic magnetism. Several previous works investigate this aspect and report the temperature dependent values of ANC in these materials. Hence, they prove to be suitable candidates for validating the \textit{C-BerryANC} code.

% For validating the code, it has been tested over three ferromagnetic materials: Fe\cite{Fe-lp}, Fe$_3$Ge\cite{drijver1976magnetic,kanematsu1963magnetic,cao2009large} and Co$_2$FeAl\cite{shukla2022atomic}. Fe is a body-centered cubic crystal while the other two materials crystallize in face-centered cubic crystal structure. Having intrinsic magnetism, these materials are expected to exhibit AHE. In this regard, several theoretical calculations and experimental measurements have been previously carried out\cite{danan1968new,dheer1967galvanomagnetic,wang2006ab,li2023anomalous,shukla2022atomic,huang2015anomalous}. These studies have verified the AHE in the materials and reported the values of AHC at various temperatures. Thus, it is convincing to use them for benchmarking the \textit{C-BerryANC} code.

\begin{figure*}[t]
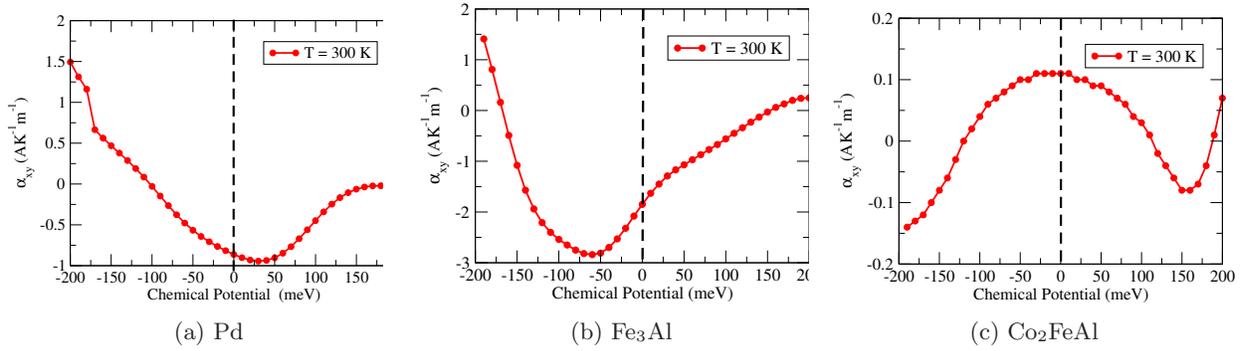

    \centering
    \begin{subfigure}[b]{0.30\textwidth}
        \includegraphics[width=\textwidth]{fig3_1.eps}
        \caption{Pd}
        \label{alphaPd}
    \end{subfigure}
    \begin{subfigure}[b]{0.30\textwidth}
        \includegraphics[width=\textwidth]{fig3_2.eps}
        \caption{Fe$_3$Al}
        \label{alphaFe3Al}
    \end{subfigure}
    \begin{subfigure}[b]{0.30\textwidth}
        \includegraphics[width=\textwidth]{fig3_3.eps}
        \caption{Co$_2$FeAl}
        \label{alphaCo2FeAl}
    \end{subfigure}
    \caption{The curve represent the $\alpha_{xy}$ vs chemical potential for respective materials at 300 K. The chemical potential is scaled with respect to Fermi energy.}
    \label{figFiles2}
\end{figure*}
 
\subsection{Computational Details}

The self-consistent ground state energy calculations for all materials are carried out using the WIEN2k package. SOC is included in these calculations. Since all the materials exhibit magnetic behavior, spin polarization is also taken into account, with the magnetization aligned along the \textit{z}-axis. Details such as the space group, lattice parameters, atomic Wyckoff positions, \textbf{\textit{k}}-mesh size, and the exchange–correlation functional used in the ground state calculations are provided in table \ref{tab3}. BZ is sampled over its irreducible wedge in all calculations. The energy convergence criterion for the self-consistent field iterations is set to 10$^{-4}$ Rydberg per unit cell.

For the ANC calculations, the \textbf{\textit{k}}-mesh size and input parameters used in the \textit{C-BerryANC} code are listed in table \ref{tab31}. The chemical potential ($\mu$) dependent values of $\alpha_{xy}$ are calculated within energy range of -200 to 200 meV relative to the Fermi level. The temperature is set at 300 K for all this computations. The $\mu$ window is divided into 40 parts. Accordingly, the parameters \textit{emin}, \textit{emax}, \textit{estep}, \textit{Tmin}, \textit{Tmax} and \textit{Tstep} are set to -200, 200, 10, 300, 300 and 25 (any non-zero value), respectively. Furthermore, for temperature-dependent calculations (within 25-300 K) at the Fermi energy, the parameters \textit{emin}, \textit{emax}, \textit{estep}, \textit{Tmin}, \textit{Tmax} and \textit{Tstep} are set to 0, 0, 1, 25, 300 and 25, respectively. All computations are performed on a system with 64 GB RAM, using \textit{NUM\_THREADS} = 28 and \textit{num\_kp\_at\_once} = 5000.

\begin{table}
\caption{\label{tab:table1}%
\normalsize{Convergence of $\alpha_{xy}$ at 300 K with respect to the size of \textbf{\textit{k}}-mesh in Fe$_3$Al. The chemical potential is set at Fermi energy.
}}
\begin{ruledtabular}
\begin{tabular}{cc}
\textrm{\textbf{\textit{k}-mesh}}&
\textrm{\textbf{$\alpha_{xy}$}}\\
\textrm{\textbf{}}&
\textrm{\textbf{$AK^{-1}m^{-1}$}}\\
\colrule
 50$\times$50$\times$50              & -1.58 \\
 100$\times$100$\times$100           & -1.87 \\
 150$\times$150$\times$150           & -1.76 \\
 200$\times$200$\times$200           & -1.84 \\
 250$\times$250$\times$250           & -1.80 \\
 300$\times$300$\times$300           & -1.79 \\
 350$\times$350$\times$350           & -1.83   \\
 400$\times$400$\times$400           & -1.83     \\
    
\end{tabular}
\end{ruledtabular}
\label{tab4}
\end{table}

\begin{table}
\caption{\label{tab:table1}%
\normalsize{Value of $\alpha_{xy}$ vs Temperature. The chemical potential is set at Fermi level.
}}
\begin{ruledtabular}
\begin{tabular}{cccc}
\textrm{\textbf{Temperature}}&
\textrm{\textbf{}}&
\textrm{\textbf{$\alpha_{xy}$}}&\\
\colrule
\textrm{\textbf{}}&
\textrm{\textbf{Pd}}&
\textrm{\textbf{Fe$_3$Al}}&
\textrm{\textbf{Co$_2$FeAl}}\\
\textrm{\textbf{(Kelvin)}}&
\textrm{\textbf{($AK^{-1}m^{-1}$})}&
\textrm{\textbf{($AK^{-1}m^{-1}$})}&
\textrm{\textbf{($AK^{-1}m^{-1}$})}\\
\colrule
 25           & -0.06  & -0.05 & 0.01 \\
 50           & -0.12  & -0.12 & 0.02 \\
 75           & -0.19  & -0.22 & 0.03  \\
 100          & -0.27  & -0.35 & 0.04   \\
 125          & -0.35  & -0.51 & 0.05  \\
 150          & -0.44  & -0.69 & 0.06    \\
 175          & -0.52  & -0.89 & 0.07   \\
 200          & -0.61  & -1.09 & 0.08  \\
 225          & -0.68  & -1.28 & 0.09  \\
 250          & -0.75  & -1.47 & 0.09  \\
 275          & -0.81  & -1.66 & 0.11 \\
 300          & -0.86  & -1.83 & 0.11   \\
    
\end{tabular}
\end{ruledtabular}
\label{tabb4}
\end{table}

\subsection{Results and discussion}
At the very first, the \textbf{\textit{k}}-point convergence of $\alpha_{xy}$ has been carried out for Fe$_3$Al. For this, the chemical potential is set at the Fermi energy while the temperature is taken to be 300 K. The obtained result is presented in table \ref{tab4}. With increase in the size of \textbf{\textit{k}}-mesh, a significant variation in the value of $\alpha_{xy}$ is observed when the size of \textbf{\textit{k}}-meshes were small. However, beyond 350$\times$350$\times$350 \textbf{\textit{k}}-mesh size, the value of $\alpha_{xy}$ gets converged to -1.83 $AK^{-1}m^{-1}$. This gives an estimate of the size of \textbf{\textit{k}}-mesh required for ANC calculations using the present code. To ensure the accuracy of the results, all further computations of ANC is performed on 400$\times$400$\times$400 size of \textbf{\textit{k}}-mesh.

The code is firstly benchmarked over Palladium (Pd). Calculations of chemical potential ($\mu$) dependent $\alpha_{xy}$ is carried out in the energy range of -200 to 200 meV around the Fermi energy. The temperature is set to 300 K for these computations. Obtained results are shown in figure \ref{alphaPd}. The Fermi energy is scaled to 0 meV. It is seen that within the given energy range, $\alpha_{xy}$ is maximum ($\sim$1.5 $AK^{-1}m^{-1}$) when the $\mu$ value is 200 meV below the Fermi level. With the rise in $\mu$, $\alpha_{xy}$ decreases and attains a minimum value of $\sim$-0.94 $AK^{-1}m^{-1}$ at $\mu$ equals to $\sim$30 meV above the Fermi energy. On further rise in value of $\mu$, $\alpha_{xy}$ is found to increase. In addition to this, the temperature dependent values of $\alpha_{xy}$ are also computed within 25-300 K, when the $\mu$ value corresponds to Fermi energy. Obtained result is shown in table \ref{tabb4}. It is found that with the rise in temperature, the magnitude of $\alpha_{xy}$ increases. At 300 K, the value of $\alpha_{xy}$ is obtained to be -0.86 $AK^{-1}m^{-1}$. Guo \textit{et. al.}, have previously studied ANC in Pd using computational approach\cite{guo2014anomalous}. In their work, the reported value of $\alpha_{xy}$ at 300 K is -0.72 $AK^{-1}m^{-1}$. Thus, the obtained result from \textit{C-BerryANC} code is found to be in a very good match with the reported data.

Nextly, using the \textit{C-BerryANC} code, the room temperature value of $\alpha_{xy}$ is calculated for Fe$_3$Al corresponding to various values of $\mu$ ranging from -200 to 200 meV around the Fermi level. Obtained results are shown in figure \ref{alphaFe3Al}. Within the given $\mu$ window, the value of $\alpha_{xy}$ is found to first decrease to attain a minimum value of -2.84 $AK^{-1}m^{-1}$ (at $\mu$ corresponding to 60 meV below the Fermi energy). With further rise in $\mu$, value of $\alpha_{xy}$ is seen to increase. In addition to this, the temperature dependent values of $\alpha_{xy}$ is computed for the material. The obtained result is shown in table \ref{tabb4}. It is found that, within 25-300K, magnitude of $\alpha_{xy}$ rises with the increase in temperature. At 300 K, the value is found to be -1.83 $AK^{-1}m^{-1}$. Sakai \textit{et. al.} have previously computed the temperature dependent value of $\alpha_{xy}$ using high-throughput computation\cite{sakai2020iron}. In their work, the maximum magnitude of $\alpha_{xy}$ for temperature values ranging upto 500 K is reported to be 2.7 $AK^{-1}m^{-1}$. For the same temperature range (\textit{i.e.}, T $\leq$ 500 K), the maximum magnitude of $\alpha_{xy}$ was reported to be 3.0 $AK^{-1}m^{-1}$ in another work\cite{koepernik2023symmetry}. In this temperature range, using \textit{C-BerryANC} code, the maximum magnitude of $\alpha_{xy}$ is found to be  2.83 $AK^{-1}m^{-1}$ (when $\mu$ is set at the Fermi energy). It is also important to highlight here that the calculations in the above mentioned works were done using tight-binding model obtained from the wannierization method. Considering this aspect, the result obtained from \textit{C-BerryANC} code is found to be in good match with the reported data.

Lastly, the \textit{C-BerryANC} code is tested over Co$_2$FeAl. As discussed above, $\mu$ dependent $\alpha_{xy}$ were computed within -200 to 200 meV around the Fermi energy and the obtained results are shown in figure \ref{alphaCo2FeAl}. It is seen that as one traces the $\mu$ values from -200 to 200 meV around the Fermi energy, value of $\alpha_{xy}$ first increases to attain a maxima ($\sim$0.11 $AK^{-1}m^{-1}$) at the Fermi energy. With further rise in the value of $\mu$, $\alpha_{xy}$ again starts to decrease to attain a minima ($\sim$-0.08 $AK^{-1}m^{-1}$) at $\mu$ equals to 160 meV above the Fermi energy. Beyond this, the value of $\alpha_{xy}$ starts to rise again with the increase in $\mu$. In addition to this, the computed values of temperature dependent $\alpha_{xy}$ within the temperature range of 25-300 K is mentioned in table \ref{tabb4}. The value of $\alpha_{xy}$ is found to increase with the rise in temperature. The value is found to be $\sim$0.11 $AK^{-1}m^{-1}$ at 300 K. Noky \textit{et. al.} have previously calculated $\alpha_{xy}$ in Co$_2$FeAl using high-throughput calculations (involving wannierization techniques) and found its magnitude to be 0.06 $AK^{-1}m^{-1}$ at 300 K\cite{noky2020giant}. The calculated value from \textit{C-BerryANC} code is found to be in fairly good match with the reported data.

It is necessary to highlight here that the calculated values of ANC from \textit{C-BerryANC} code may sometime deviate largely from the available emperimental results. The possible reason for this may be defect or impurity in the sample. It may also arise if the sample is in polycrystalline phase. These conditions lead to various scattering mechanisms which is expected to be the cause for such deviations. Thus, user is suggested to consider these aspects while comparing the calculated results with the experimental data.

\section{Conclusion}
In this work, we have developed \textit{C-BerryANC}, a robust and efficient code for calculating the Berry-curvature ($\boldsymbol\Omega$) driven anomalous Nernst conductivity (ANC) from \textit{first-principles} data. The code uses eigenvalues and momentum-matrices elements obtained from WIEN2k calculations and employs the Kubo formula to compute the $\boldsymbol\Omega$ across the Brillouin zone (BZ). By incorporating the Fermi-Dirac distribution, \textit{C-BerryANC} effectively accounts for the dependence of electronic occupation on temperature and chemical potential, allowing for realistic simulation of ANC in topological materials. Furthermore, the implementation of parallel computation for both $\boldsymbol\Omega$ \& ANC, combined with the efficient storage and management of band-resolved $\boldsymbol\Omega$ in binary format, significantly improves the performance and user-friendliness of the code. For benchmarking of the code, it has been tested on some well-studied ferromagnetic materials exhibiting ANC. These include- Pd, Fe$_3$Al \& Co$_2$FeAl. The values of ANC of these materials, calculated from \textit{C-BerryANC} code, are found to be in good match with those reported in literature. For instance, the room temperature value of $\alpha_{xy}$ for Pd is obtained to be -0.86 $AK^{-1}m^{-1}$ which is in good match with the previously reported data (-0.72 $AK^{-1}m^{-1}$\cite{guo2014anomalous}). In case of Fe$_3$Al, the maximum value of $\alpha_{xy}$ for $T\leq$500 K is computed as 2.83 $AK^{-1}m^{-1}$ which is also in good agreement with previously reported works (3.0 $AK^{-1}m^{-1}$\cite{koepernik2023symmetry},2.7 $AK^{-1}m^{-1}$\cite{sakai2020iron}). Lastly, for Co$_2$FeAl, the value of $\alpha_{xy}$ is obtained to be $\sim$0.11 $AK^{-1}m^{-1}$ at 300 K which is in fairly good match with data available in literature (0.06 $AK^{-1}m^{-1}$ at 300 K\cite{noky2020giant}). These findings confirm the precision, computational efficiency, and robustness of the \textit{C-BerryANC} code in evaluating the ANC for a wide range of materials.

\bibliographystyle{apsrev4-2}
\bibliography{references} % references.bib should contain your bibliography

\end{document}